\documentclass[12pt]{article} 
\usepackage{amsmath,amssymb,amsfonts}
\usepackage{hyperref}
\usepackage[usenames]{color}
\usepackage{graphicx}
\usepackage{lineno}
\newcommand{\bc}{\begin{center}}
\newcommand{\ec}{\end{center}}
\newcommand{\be}{\begin{equation}}
\newcommand{\ee}{\end{equation}}
\newcommand{\bea}{\begin{eqnarray}}
\newcommand{\eea}{\end{eqnarray}}
\newcommand{\bef}{\begin{figure}}
\newcommand{\ef}{\end{figure}}
\def\mean#1{\left< #1 \right>}
\newcommand{\mub}{\mu_{\scriptscriptstyle B}}

\newcommand{\snn}{\sqrt{s_{\scriptscriptstyle NN}}}

\newcommand{\ie}{{\sl i.e.\/}}

\title{Freeze-out and thermalization\\ in relativistic heavy ion collisions}
\author
{Sourendu Gupta,$^{1}$ Debasish Mallick,$^{2}$ Dipak Kumar
  Mishra,$^{3}$, \\ Bedangadas Mohanty,$^{2,\footnote{Chinese Academy
      of Sciences President's International Fellowship Initiative
      visiting scientist at Institute of Modern Physics, Lanzhou 730000, China}}$, Nu Xu$^{4,5}$\\
\\
\normalsize{$^{1}$Department of Theoretical Physics, Tata Institute of Fundamental Research,}\\
\normalsize{Homi Bhabha Road, Mumbai 400005, India}\\
\normalsize{$^{2}$School of Physical Sciences, National Institute of
  Science Education and Research, HBNI,}\\
\normalsize{Jatni 752050, India}\\
\normalsize{$^{3}$Nuclear Physics Division, Bhabha Atomic Research
  Centre, Mumbai 400085, India,} \\
\normalsize{$^{4}$Institute of Modern Physics, Chinese Academy of
  Sciences,}\\
\normalsize{509 Nanchang Road, Lanzhou 730000, China}\\
\normalsize{$^{5}$College of Physical Science and Technology, Central China Normal University,} \\
\normalsize{152 Luyu Road, Wuhan 430079, China}\\
}
\begin{document}

\baselineskip24pt

\maketitle
\begin{abstract}
 High energy heavy-ion collisions in laboratory produce a form
of matter that can test Quantum Chromodynamics (QCD), the theory of
strong interactions, at high temperatures. One of the exciting
possibilities is the existence of thermodynamically distinct states
of QCD, particularly a phase of de-confined quarks and gluons.
An important step in establishing this new state of QCD is to demonstrate
that the system has attained thermal equilibrium. We present a test of thermal
equilibrium by checking that the mean hadron yields produced in the
small impact parameter collisions as well as grand canonical
fluctuations of conserved quantities give consistent temperature and baryon
chemical potential for the last scattering surface.  This consistency for moments up to
third order of the net-baryon number, charge, and strangeness is a key step in the proof
that the QCD matter produced in heavy-ion collision attains thermal
equilibrium. It is a clear indication for the first time, using
fluctuation observables, that a
femto-scale system attains thermalization. The study also indicates that
the relaxation time scales for the system are  
comparable to or smaller than the life time of the fireball.

\end{abstract}

Relativistic collisions of heavy-ion are being carried out at the
Relativistic Heavy Ion Collider (RHIC) in BNL and the Large Hadron
Collider (LHC) facilities in CERN. These experiments have demonstrated the existence of a deconfined
state of quarks and gluons, which are the fundamental constituents
of baryonic matter. The experiments observe the collective flow of a
fireball made of such deconfined matter \cite{ white3,
white4, Aamodt:2010pa, Aamodt:2010jd, Chatrchyan:2011sx}
(see also a review \cite{Gyulassy:2004zy}). This state of matter replicates baryonic matter in the universe when
it was a few microseconds old. It is also conjectured that the cores
of astrophysical objects like neutron stars may contain a related form
of baryonic matter, albeit at lower temperature but higher density
\cite{Lattimer:2004pg}.

By studying the flow of matter in the femto-scale fireballs produced
at RHIC and LHC, it has been shown that the ratio of shear viscosity
to entropy density, $\eta/s$, is the lowest among all known fluids
\cite{Csernai:2006zz,Romatschke:2007mq}. The values of this ratio extracted from experiment
are close to the bounds expected for a strongly coupled quantum fluid
\cite{Kovtun:2004de}.  Further, it has been recently shown this fluid
has large vortical effect \cite{STAR:2017ckg}.

Theoretical studies of strongly interacting matter using lattice field
theory shows a cross over from hadronic matter to a deconfined state of
quarks and gluons \cite{Aoki:2006we}. Consistent with these expectations,
no sign of a phase transition has been found at the top energies in the
experiments at RHIC and LHC.

Lattice studies have revealed an experimentally reachable critical
point
\cite{Gavai:2004sd, Datta:2016ukp, DElia:2018fjp}  in an extremely rich phase
diagram of strongly interacting matter \cite{Stephanov:1999zu,
Hatta:2003wn, Stephanov:2008qz,Asakawa:2009aj}.
Currently experimental studies are underway to vary the collision
energy in order to study this phase diagram \cite{Adamczyk:2017iwn,BraunMunzinger:2008tz, Fukushima:2010bq}.
Experimental data reveals tantalising hints of critical behaviour
\cite{Adamczyk:2013dal, Aggarwal:2010wy, Luo:2015ewa},  but these
interpretations of experiments have the underlying assumption
that the fireball produced in the collisions should have come to
local thermal equilibrium during its evolution.  Experimental tests of
thermalization are non-trivial for these femto-scale system: not just
because the systems are small, but also because they are expanding~\cite{Adamczyk:2017iwn}.
This is what we examine critically in this paper.

Systems which are significantly larger than the measuring instruments
allow ensembles of measurements on a single system, and one may
exploit the hierarchy of scales between the size of the full system, the
measuring apparatus, and the microscopic length scales. This allows us to
create ensembles of measurements on identical systems by measuring the
system at multiple points. This philosophy is used routinely to measure
fluctuations in the early universe \cite{Smoot:1992td, Komatsu:2010fb}. On
the other hand it is problematic when the measurement involves the whole
system. This happens for low-multipole fluctuations in the universe, and
is called the cosmic variance problem. For the femto-scale system which
we study, the hierarchy of length scales collapses for all measurements,
and one constructs the ensemble by repeating the experiment. This
requires additional checks to make sure that the repeated trials are
indeed comparable.

Another dimension of the problem is whether the system evolves with time
or not. Again when the time scale of evolution, of measurement, and the
microscopic dynamics are well separated, simplicities emerge.  In a static
large system, repeated measurements can be made at effectively the same
time. This can probe thermodynamic stability such as the convexity of free
energies, or the lack of entropy production. However, in quasi-statically
evolving systems the entropy may change simply because the temperature
changes with time, and not because the system is driven towards or out
of equilibrium.  When there is no clear hierarchy between the rate of
evolution and microscopic relaxation time, then one has to qualify the
notion of thermal equilibrium.

Distribution functions are driven out of equilibrium, and different parts
may evolve at different rates. For a small decrement of temperature,
$\Delta T$, the equilibration of mean energies will require small changes.
In a Boltzmann gas, for example, the average energy changes by the factor
$3\Delta T/2$. This change is easily made by small exchanges of momentum
among those particles which lie near the peak of the thermal distribution,
and can proceed quickly.
The tail of the distribution, on the other 
hand, has to change by a large factor.
There are
two ways for this to happen.  One is that particles carrying these rare
large momenta collide with others having equally large momenta and both
emerge with smaller momentum in one scattering. Another way is for the
large momentum particle to collide many times with particles of much
smaller momenta and decrease the momentum in small steps. Either route
gives longer relaxation times. Similarly, local changes in the chemical
potential, $\Delta\mu$, distort the local number distribution, different
part of which relax at different rates.  Due to such differential
relaxation of different parts of the distribution function different
moments of the distributions may not correspond to the same temperature
and chemical potential. One also knows that approach to a critical
point brings with it correlated fluctuations of the whole system in both
space and time, thereby increasing relaxation times
dramatically. Studying the moments of the number distributions to find evidences
for both equilibrium and large departures scenarios are therefore crucial tests for the
physics of these femto-systems.

In the relativistic femto-systems created by heavy-ion collisions,
the conserved quantities are the net-baryon number ($B$), strangeness
($S$) and electric charge ($Q$). Since these are additive quantum numbers, every
anti-particle has a charge opposite to that of the particle, and the net
quantum number is the algebraic sum over particles and anti-particles.
Conserved quantities measured for the whole system will always be equal
to that in the initial state. However, using the limited acceptance of the
detector, only a part of the thermalized system is accessed experimentally
and hence can be treated within the framework of grand canonical ensemble.
This allows us to measure fluctuations of the conserved quantities.

If the fireballs were static and in thermal equilibrium, they could
be described by a temperature, $T$, and three chemical potentials
$\mu_B$, $\mu_S$ and $\mu_Q$. Since the system is not static, these
four quantities are functions of space and time. One can only observe
particles which come to the detector after the last scattering. This
region of last scattering is characterized by a specific value of $T$
and chemical potentials which are generally called freezeout parameters.
It is common to quote as the freezeout parameters the values of these
thermodynamic variables which give the best match between particle yields
(i.e means of number distributions) in a statistical model of an ideal gas
of hadrons and resonances (HRG)\cite{Cleymans:2005xv, Andronic:2005yp,
Becattini:2003wp, Garg:2013ata, Andronic:2017pug}. Characterizing the yield at fixed
$\snn$ and acceptance window by a single set of freezeout parameters
was seen to work well for charged hadrons \cite{Cleymans:2005xv,
Andronic:2005yp}. Characterizing strange and non-strange hadrons by
different freezeout conditions works slightly better in some cases
\cite{Chatterjee:2013yga}. Other variants of the HRG have also been
shown to work \cite{Venugopalan:1992hy, Yen:1997rv, Chatterjee:2009km,
  Vovchenko:2016rkn, Samanta:2017yhh, Chatterjee:2017yhp}.
However, these HRG models also seem to describe yield data in 
$e^+e^-$ and pp collisions, where one does not expect thermalized 
matter to be formed \cite{Becattini:1995if, Becattini:1997rv, 
Das:2016muc}. Similarly, they can describe yields in highly peripheral 
collisions. These observations introduce a severe uncertainty in the 
interpretation of the freezeout parameters derived using yields in terms 
of thermal conditions. Here we propose that a way out of this impasse 
is to ask for common thermal descriptions of more than just the mean 
particle number.

{\bf Model:} The hadron resonance gas (HRG) model that we utilize in this work is
briefly discussed below and details can be found at Ref.\cite{Garg:2013ata}.
In a grand canonical ensemble with a gas of hadrons
and resonances, the thermodynamic pressure ($P$) can be obtained
from the logarithm of partition function in the limit of large volume as
\be
 P(T,\mu_B,\mu_Q,\mu_S,V) =\frac TV\sum_i \ln Z_i= \sum_i
   \pm\frac{T g_i}{2\pi^2}\int d^3k\ln
     \left\{1\pm\exp\left[(\mu_i-E)/T\right]\right\}.
\ee
where $k$ is the momentum, $g_i$ is the degeneracy factor of the $i^{\rm
th\/}$ species of hadron or resonance, and the plus signs are for Fermions
whereas the minus signs are for Bosons.  The sum is carried out over
all particles in thermal equilibrium with masses up to 2.5 GeV which are listed in Particle
Data Group (PDG) booklet.

The freezeout conditions for the fireball are characterized by the
temperature $T$, freezeout volume $V$, and the three chemical potentials $\mu_B$, $\mu_Q$,
and $\mu_S$.  If the system is in chemical equilibrium then the chemical
potential for the $i^{\rm th\/}$ species is given by $\mu_i = B_i\mu_B + Q_i\mu_Q
+ S_i\mu_S$, where $B_i$, $Q_i$ and $S_i$ are the baryon number, electric
charge  and strangeness of the particle.
In the analysis that we present in this paper, the detector's
acceptance and resonance decay effects are taken
into account, for details see Ref.\cite{Garg:2013ata}.

The effect of radial flow (collective expansion of the system) seen in heavy-ion collisions do not affect
the moments as they are obtained integrated over the measured $p_{T}$
range. HRG model calculations with and without flow shows a small
difference for the observable studied~\cite{Garg:2013ata} and those
are used as systematic uncertainties associated with the model
calculations in the paper.

The grand canonical approach is justifiable as experimental data used
in the current work analyses a portion of the fireball produced in
heavy-ion collisions. It is akin to an open system, unlike for the
data for full 4$\pi$ coverage which would have required a canonical
treatment with exact conservation of charges. In addition, the
criteria  for applicability of grand canonical ensemble, $VT^3>1$
holds true for the bulk of the produced particles in heavy-ion
collisions. Such an approach has been widely used to understand the yields of
produced hadrons~\cite{Adamczyk:2017iwn,BraunMunzinger:2007zz,
  Becattini:2012xb} and
fluctuations~\cite{Gavai:2010zn, Bazavov:2012vg,Borsanyi:2014ewa}  in the field of
high-energy nuclear collisions.
Further, as discussed in Ref.~\cite{Bzdak:2012an}, the effect of global
conservation of charges depends on the fraction of charges accepted in
the detector, which is found to be small and within the experimental
uncertainties for the data up to third order moment used in the current work.

{\bf Observable:} In this paper, 
we study the cumulants of different order such as:
$C^X_1=\mean X$, $C^X_2=\mean {(X-\mean X)^2}$, $C^X_3=\mean {(X-\mean
X)^3}$, $C^X_4=\mean {(X-\mean X)^4} - 3(C^X_2)^2$ and so on, where $X$
can be one of $B$, $Q$ and $S$.  Since the observed cumulants in thermal
equilibrium are related to the susceptibilities:
\be
  C^X_n = (V/T)T^n\chi^{(n)}_X\left(T,\mub\right),
\label{cum}\ee
there is a possible way to check thermalization and the predictions of
lattice QCD
\cite{Bazavov:2012vg,Gupta:2009mu,Gavai:2010zn,Gupta:2011wh,Cheng:2008zh,Bellwied:2013cta}. The
$n^{\rm th\/}$ order generalised susceptibilities ($\chi_X^{(n)}$), where $X$
represents baryon, strangeness or electric charge indices, can be 
expressed as \cite{Gavai:2004sd}, 
\be 
  \chi_X^{(n)}(T,\mu_B,\mu_Q,\mu_S) =
    \frac{d^n P(T,\mu_B,\mu_Q,\mu_S)}{d\mu_X^n}. 
\ee

Usually ratios of cumulant or products
of moments are calculated in order to cancel the dependence on system
volume. Experiments use the notation $M=C_1$ for the mean, $\sigma^2=C_2$
for the variance, $S=C_3 /C_2^{3/2}$ for the skewness, and
$\kappa=C_4/C_2^2$ for the kurtosis. In terms of these variables, the ratios of cumulants
are $\sigma^2/M=C_2/C_1$, $S\sigma=C_3/C_2$ and $\kappa\sigma^2=C_4/C_2$~\cite{Adamczyk:2013dal, Adamczyk:2017wsl, Adamczyk:2014fia}.
Though all the three ratios of higher cumulants ($\sigma^2/M$, $S\sigma$,
$\kappa\sigma^2$) and second order off-diagonal cumulants
($\sigma^{2}_{XY}$)~\cite{Adam:2019xmk} reported by the STAR experiment are expected to carry
the fluctuation signals, only observables up to third order
moments  ($M$, $\sigma^{2}_{XY}$, $\sigma^{2}_{XX}$, $S\sigma$) 
are used in this work. The observable, $\kappa\sigma^2$, will be used
elsewhere to test
departures from thermal equilibrium. The observables calculated in the
HRG model are the same as measured in the experiment.

{\bf Experimental Data:} The STAR experiment has published measurements on
cumulants up to 4th order and their ratios for the net-proton number (NP) \cite{Adamczyk:2013dal},
the net-kaon number (NK) \cite{Adamczyk:2017wsl} and the net-charge
(NQ) \cite{Adamczyk:2014fia}. The experiment has also published second
order off-diagonal cumulants~\cite{Adam:2019xmk}. The effect of choosing NP as a proxy of net-baryon and NK as
a proxy of net-strangeness has been studied in detail in
\cite{Kitazawa:2012at,Zhou:2017jfk}. Event-by-event net-charge
distributions are obtained using the Time Projection Chamber (TPC)
and Time of Flight (TOF) detectors at STAR in the momentum range
$0.2 < p_T (GeV/c) < 2.0$ and a pseudo-rapidity range of $|\eta| <
0.5$ \cite{Adamczyk:2014fia}. Net-proton distribution is measured
\cite{Adamczyk:2013dal} using TPC detector only in a momentum range
of $0.4 < p_{T} (GeV/c) < 0.8$ and net-kaon \cite{Adamczyk:2017wsl}
using both TPC and TOF detector within a momentum range of $0.2 <
p_{T} (GeV/c) < 1.6$. The second order off-diagonal cumulants
$\sigma^{2}_{Qp}$, $\sigma^{2}_{QK}$ and $\sigma^{2}_{pK}$ representing
correlations between electric charge-baryon number,
electric charge-strangeness number and baryon-strangeness number are
obtained by STAR by detecting particles within of acceptance of
$|\eta| < 0.5$ and $0.4 < p_{T} (GeV/c) < 1.6$~\cite{Adam:2019xmk}.
All the above measurements are done for
small impact parameter (called central 0-5\%) Au on Au collisions at a
range $\snn=$ 7.7--200 GeV.  The cumulants of distributions are corrected
for detector inefficiencies and other analysis artefacts as discussed in
\cite{Adamczyk:2013dal, Adamczyk:2017wsl, Adamczyk:2014fia}.  So the
experimental data available for this work are the observables, $M_{\pi^{+}}$, $M_{\pi^{-}}$,
$M_{K^{+}}$, $M_{K^{-}}$, $M_{p}$, $M_{\bar{p}}$,
$\sigma^2(NP)$,$\sigma^2(NK)$, $\sigma^2(NQ)$, $\sigma^{2}_{Qp}$,
$\sigma^{2}_{QK}$ and $\sigma^{2}_{pK}$, $S\sigma(NP)$, $S\sigma(NK)$ and $S\sigma(NQ)$.
It has been previously shown by several authors 
\cite{Mishra:2016qyj,Nahrgang:2014fza,Alba:2014eba} that the $\sigma^2/M$ for net-charge and net-protons are not properly explained by several variants 
of HRG models. The reason has been understood to be due to the acceptance 
limitation in the measurements --- resonance decays can throw part of the charge outside 
the detector acceptance, thereby increasing the value of $\sigma^2/M$
over the intrinsic thermal distribution for the
fireball. Interestingly the decay distortion is absent from 
the observables $\sigma^{2}/M(NK)$ and all three $S\sigma$. 
  In order to deal with the stochasticity due to decays, we introduce
  the three volume independent observable using the second order cumulants for the current study:
$\sigma^{2}_{QK}$/$\sigma^2(NK)$, $\sigma^{2}_{pK}$/$\sigma^2(NK)$,
and $\sigma^{2}_{Qp}$/$\sigma^2(NP)$~\cite{Gavai:2005yk}.

{\bf Methodology:} The freezeout parameters, namely the $T$, $\mu_B$, $\mu_S$ and $\mu_Q$, are extracted at
each $\snn$, from the experimental data by minimizing
\be
 \chi^2=\sum_{i=1}^N\left(\frac{\Delta_i}{E_i}\right)^2
   \qquad{\rm where}\qquad
   \Delta_i=R_i^{\rm exp\/}-R_i^{\rm HRG\/},
\label{chisq}\ee
where $N$ is the number of observables used in this calculation. $R_i^{\rm
exp\/}$ and $R_i^{\rm HRG\/}$ are experimental measurements and HRG
model calculations respectively for the $i^{\rm th\/}$  observable, and $E_i$ is
the statistical uncertainty in the experiment.

As a first step in the analysis we tested whether a fit can distinguish
between a thermal and a non-thermal system. This was done by taking
simulated data obtained from the non-thermal model called UrQMD
\cite{Bleicher:1999xi}, and trying to extract freezeout conditions from
it by the process outlined in Eq.~\ref{chisq}. Approximately one million UrQMD
events were analyzed for 0-5\% Au+Au collisions at each collision
energy. In this case we found that
the value of $\chi^2$/ndf $\sim$ $O(1000-10000)$ for all observable rules out a reasonable fit, and the best possible
freezeout parameters obtained from the yields, $\sigma^{2}/M(NP,
NK,NQ)$, and $S\sigma(NP,NK,NQ)$ are quite different. Even carrying out
the studies without $\sigma^2/M$ for net-charge and net-protons  in
UrQMD yields a $\chi^2$/ndf $\sim$ $O(100-10000)$. Since UrQMD model is
known to reproduce many other aspects of the final state obtained in heavy-ion
collisions like multiplicity and mean $p_{T}$~\cite{Petersen:2008kb}, this shows that accidental agreement of the data with the
HRG model is unlikely. We have additionally repeated the analysis using a different
non-thermal model that includes parton degrees of freedom, A Multi Phase Transport (AMPT) model~\cite{Lin:2004en}, for Au+Au
collisions at $\snn$ = 200 GeV and found that the thermal fit failed.

\bef[tbh!]\bc\includegraphics[scale=0.65]{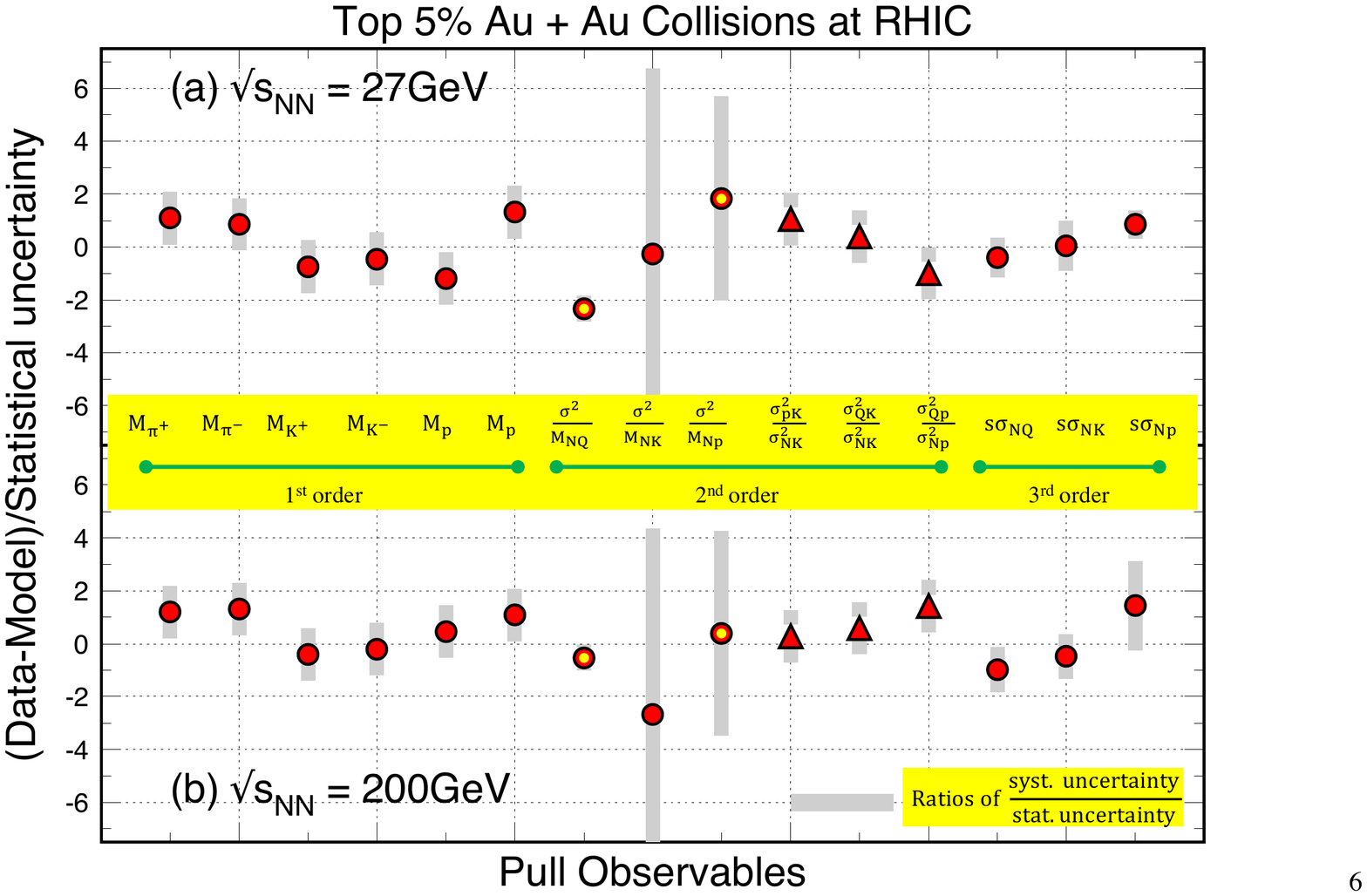}\ec
\caption{Detailed comparison of the best fit predictions of the HRG model
at (a) $\snn=27$ GeV and (b) $\snn=200$ GeV, which minimizes $\chi^2$ given in Eq.~(\ref{chisq})
with the experimental measurements. This representative case at two
collision energies shows the difference between data and
HRG model ($\Delta$) divided by the statistical uncertainty along the ordinate for each of 
the observables mentioned in the absicissa.
The $\sigma^2/M$(NQ) and $\sigma^2/M$(NP) values
  shown as open circles are scaled down by a factor 100. The figure also shows a comparison of the
magnitudes of systematic and statistical uncertainties, since the former are not
included in the definition of $\chi^2$ to improve the discriminatory
power of the fits.}
\label{basic}\ef

{\bf Results:} In Figure \ref{basic} we show the quality of a fit to fifteen pieces of
data (shown in x-axis) at $\snn=27$ GeV and $\snn=200$ GeV obtained by varying only $T$,
$\mu_B$, $\mu_S$ and $\mu_S$. The $\sigma^2/M$ for net-charge and net-protons behaves quite
differently from the others due to the effect of resonance decay as
discussed above. However, the ratios of second order cumulants
($\sigma^{2}_{QK}$/$\sigma^2(NK)$, $\sigma^{2}_{pK}$/$\sigma^2(NK)$, and
$\sigma^{2}_{Qp}$/$\sigma^2(NP)$) are well explained by the model.

\bef[tbh!]\bc\includegraphics[scale=0.7]{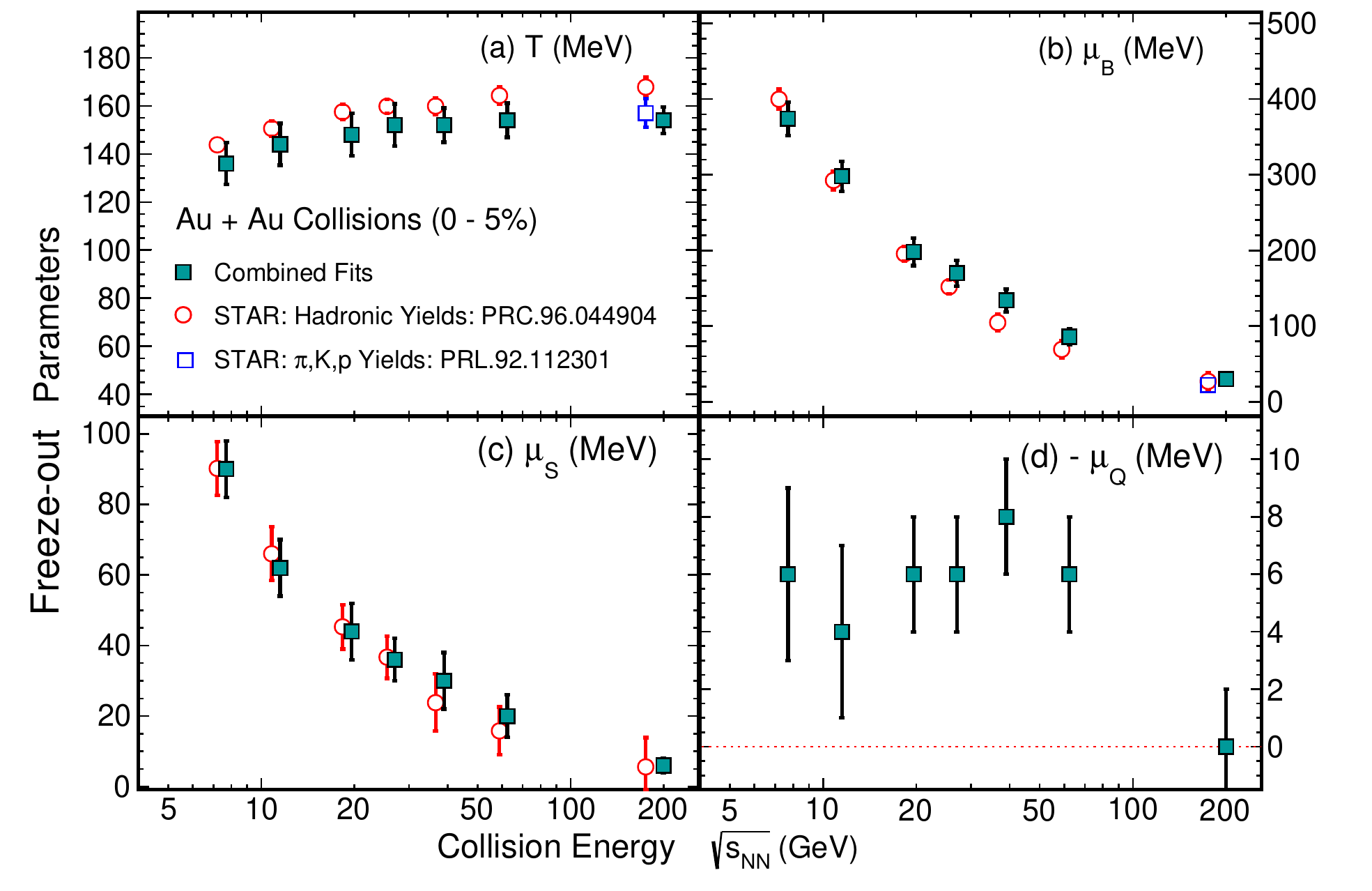}\ec 
 \caption{The best fit parameters of the HRG model at different $\snn$
   obtained by fitting data for central Au+Au collisions. Shown as 
   red open circles (with points slightly displaced for clarity of presentation) 
   are the comparison of the freezeout parameters to those 
obtained using only the mean yields of various produced hadrons by STAR 
experiment \cite{Adamczyk:2017iwn}. Also shown as open blue squares are the comparison of the freeze-out parameters 
at $\snn=200$ GeV obtained by fitting to the mean yields of only $\pi^{\pm}$, $K^{\pm}$ and $p(\bar{p})$.}
\label{thermal-parameters}\ef 
At every $\snn$, we fitted the parameters of the HRG model to the 
thirteen observables ($M_{\pi^{+}}$, $M_{\pi^{-}}$,
$M_{K^{+}}$, $M_{K^{-}}$, $M_{p}$, $M_{\bar{p}}$, $\sigma^2/M(NK)$,$\sigma^{2}_{QK}$/$\sigma^2(NK)$, $\sigma^{2}_{pK}$/$\sigma^2(NK)$,$\sigma^{2}_{Qp}$/$\sigma^2(NP)$, $S\sigma(NP)$,$S\sigma(NK)$ and $S\sigma(NQ)$) shown in Figure \ref{basic} as solid circles. In Fig.
\ref{thermal-parameters} we show the best fit values of $T$, $\mu_B$,
$\mu_S$ and $\mu_Q$ as a function of $\snn$. The $\chi^2$/ndf are of the
$O(1)$, except for $\snn=7.7$ GeV, where the value is 11.9. The results of the fits, which now include higher second and third order
moments of particle multiplicity distributions are in good agreement
with those obtained only using the mean yields of produced particles
(shown as red open circles) by STAR experiment \cite{Adamczyk:2017iwn}. The slight difference in $T$ values
arises because of inclusion of multi-strange hadrons in the fits to
the yields. As shown in the figure, for $\snn=200$ GeV (as blue open
square), using similar observables as in the current study, i.e the yields of $\pi^{\pm}$, $K^{\pm}$ and $p(\bar{p})$,  has
a better agreement with our results. Including systematic
uncertainties in the experimental data by adding in quadrature improves the $\chi^2$/ndf
values to 1.0 and 4.6 for $\snn = 200$ and 7.7 GeV, respectively.
We have verified that the
measured $p_{T}$ distributions of pion, kaon, proton and the
anti-particles for 0-5\% Au+Au collisions at $\snn$ = 200 and 19.6 GeV
are reproduced using a thermal model with the extracted thermal
parameters and average radial flow velocity of 0.55$c$ and 0.46$c$, respectively.

\bef[tbh!]\bc\includegraphics[scale=0.7]{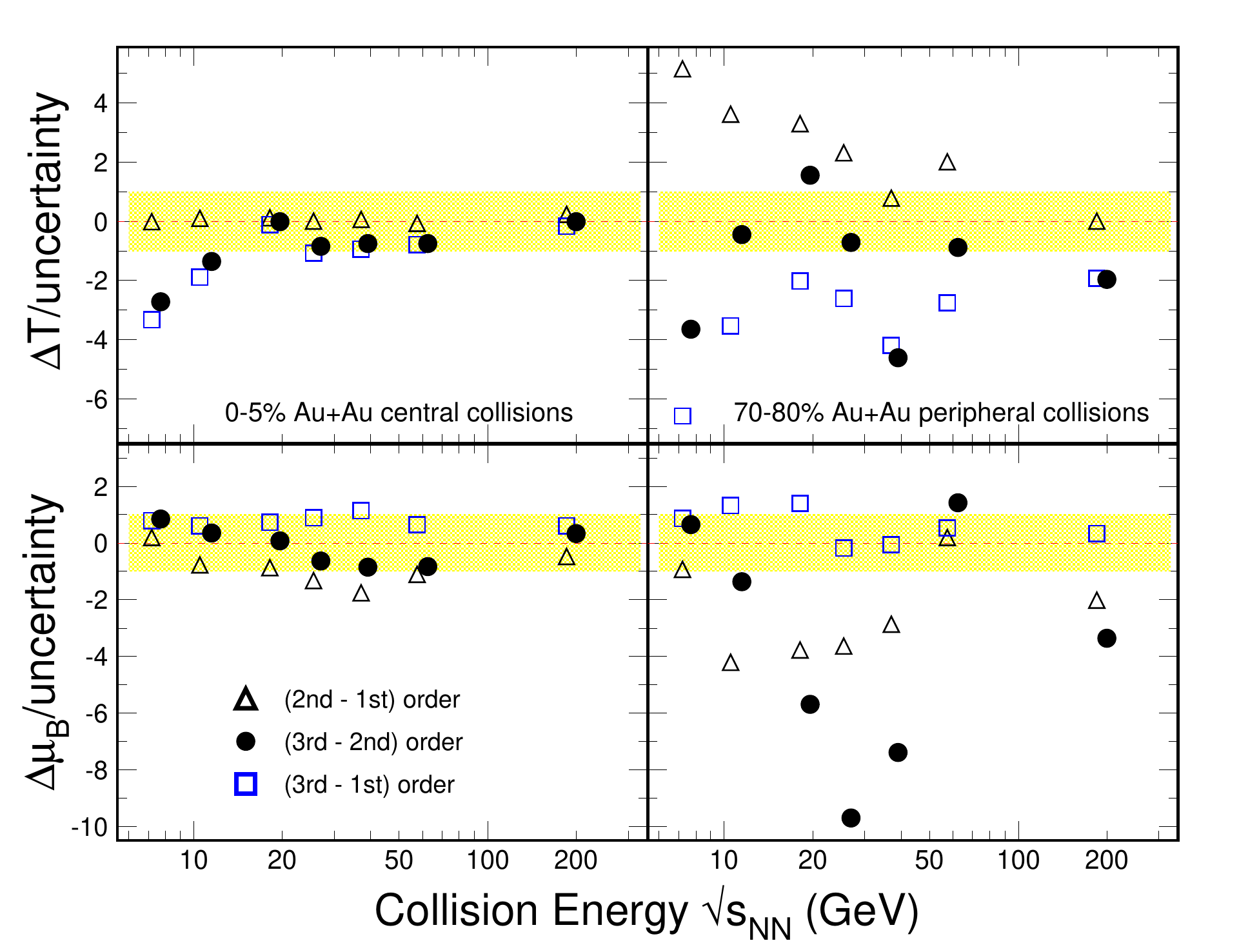}\ec 
 \caption{Detailed comparison of the best fit parameters of the HRG model 
at different $\snn$ obtained by fitting to different subsets of the
data. The difference in temperature ($\Delta T$) and baryon chemical
potential ($\Delta\mu_B$) from the third order to
  second order, third order to first
order and second order to the first order moments are shown as
filled-circles, open-squares and open-triangles, respectively.
The figures show that consistent values of the freezeout parameters are obtained 
from different subsets of the data in central collisions at all $\snn$
except, possibly, the two smallest values. In peripheral collisions 
fits to different subsets of data give significantly different results, 
implying that thermalization is not seen.}
\label{deviation-obyo}\ef

In Figure \ref{deviation-obyo} we show the $\Delta T$ and
$\Delta\mu_B$, the differences
between the best fit value of $T$ and $\mu_B$ obtained from
yields of $\pi^{\pm}$, $K^{\pm}$ and $p(\bar{p})$, 
second order cumulants
$\sigma^2/M$(NK), $\sigma^{2}_{QK}$/$\sigma^2(NK)$,
  $\sigma^{2}_{pK}$/$\sigma^2(NK)$,
  $\sigma^{2}_{Qp}$/$\sigma^2(NP)$ and those obtained by fitting the
three $S\sigma(NP)$,  $S\sigma(NK)$ and $S\sigma(NQ)$ for various $\snn$. Results for both
central (left panel) and peripheral (right panel) collisions are shown. Since the minimum number of
observables in a set is three, we have kept $T$ and $\mu_B$ parameters
of HRG model free, fixed $\mu_S$ values to those from STAR
experiment \cite{Adamczyk:2017iwn} and $\mu_{Q}$ values to zero, for
this test only. We note that in central collisions for
$\snn=19.6$ GeV and higher collision energies, the best fit freezeout parameters obtained from
different observables agree with each other within uncertainties. We further
note that in peripheral collisions the different observables give
very different freezeout conditions. A similar study with UrQMD
central Au+Au collision data yields $\Delta T$ and $\Delta\mu_B$ values normalized to the
respective uncertainties that varies in the range --12 to 18, indicating no
agreement between HRG thermal parameter values from the UrQMD data.
The fact that results from HRG model are in excellent agreement
with experimental data on various orders of fluctuation measure and with
similar values of extracted thermal parameters $T$ and $\mu_{B}$
within the uncertainties, demonstrates that the QCD matter produced in
central Au on Au collisions at RHIC has attained thermalization at
least for $\snn=19.6$ GeV and above.

\bef[tbh!]\bc\includegraphics[scale=0.6]{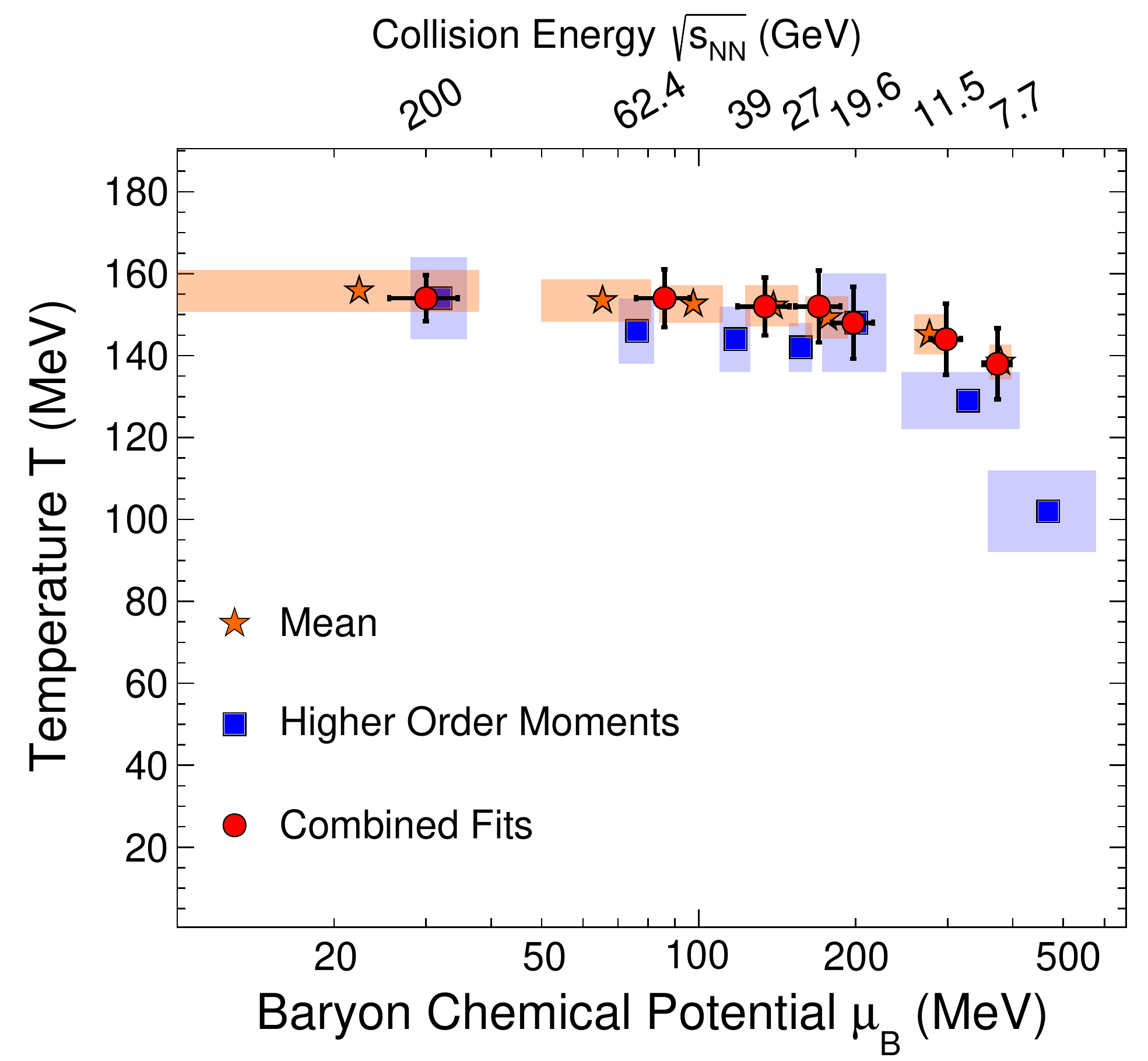}\ec
\caption{T and $\mu_B$ values at freeze-out for Au on Au collision
(0--5\% centrality) at $\snn=$ 7.7--200 GeV. The circles represent the
best fit T and $\mu_B$ values (black lines the uncertainties) obtained from the comparison
of experimental yields of pion, kaon and proton, $\sigma^{2}/M$ of
net-kaon, $\sigma_{QK}$/$\sigma^2(NK)$, $\sigma_{pK}$/$\sigma^2(NK)$,
$\sigma_{Qp}$/$\sigma^2(NP)$ and $S\sigma$ of net-proton, net-kaon and net-charge to corresponding results
from HRG model with all four parameters free. Square markers represent T and $\mu_B$ values
(blue band the uncertainties) obtained from the simultaneous
comparison of 2nd-order and 3rd-order moments to corresponding
results from the HRG model.
The stars are the results corresponding to the mean values
of the yields of pion, kaon and protons. Coloured bands show uncertainties on freeze-out $T$
and $\mu_B$ values.}
\label{final}\ef

Our final results are shown in Figure \ref{final}. This shows the results 
of the extraction of the freezeout parameters simultaneously using the 13 
observables discussed earlier at each $\snn$ GeV. Also shown are the freeze-out $T$ and $\mu_B$
extracted from data and the HRG model using the combination of $M_\pi$,
$M_K$ and $M_p$, 
and combined higher order moments of $\sigma^{2}/M$ of
net-kaon, $\sigma^{2}_{QK}$/$\sigma^2(NK)$, $\sigma^{2}_{pK}$/$\sigma^2(NK)$,
$\sigma^{2}_{Qp}$/$\sigma^2(NP)$, $S\sigma(NP,NK,NQ)$ observable for Au on Au collisions at $\snn=$
7.7--200 GeV. The uncertainties are systematic associated with the implementation of various assumptions in the
model and statistical, both added in quadrature. The model assumptions
varied includes, radial flow effect, interactions through excluded
volume effect, fixing some of the model parameters like $\mu_{Q}$ and
$\mu_{S}$ from other measurements, resonance decay and considering width of resonances. The extracted $T$ and $\mu_B$ values are in agreement among
all the three observables at each $\snn$ within the uncertainties upto
$\snn=19.6$ GeV, below that there are clear indications of deviations.
The agreement between freezeout parameters extracted from different
pieces of data shows that at freezeout the system is thermalized at
large $\snn$.

{\bf Conclusions:}

We have reported a study of thermalization at the last
stages of the evolution of the fireball created in heavy-ion collisions.
The analysis of yields of hadrons (\ie, the means of the hadron number
distributions) is known to be explained in thermal models for central
and peripheral collisions of nuclei, in pp and $e^+e^-$ collisions, and
for synthetic data produced through event generators like UrQMD. This
raises doubts about whether the analysis of yields is sufficient to tell
us about thermalization of the fireball.

We found that a sensitive test of thermalization is the simultaneous
description of yields as well as the event-to-event fluctuation
measures $\sigma^{2}/M$ (for net-kaons), $\sigma^{2}/M$, $\sigma^{2}_{QK}$/$\sigma^2(NK)$, $\sigma^{2}_{pK}$/$\sigma^2(NK)$,
$\sigma^{2}_{Qp}$/$\sigma^2(NP)$ and $S\sigma$ (for net-charge,
net-protons and net-kaons).
The variance measure $\sigma^{2}/M$ for net-charge and net-protons has
been found to be sensitive to acceptance, due to resonance decays. As a
result, we cannot use these measures along with the
rest. Instead we use the observables,
  $\sigma^{2}_{QK}$/$\sigma^2(NK)$, $\sigma^{2}_{pK}$/$\sigma^2(NK)$, $\sigma^{2}_{Qp}$/$\sigma^2(NP)$.
Agreement of all these observables for
central Au+Au collisions with
thermal gas predictions under common freezeout conditions is a strong
indication that not only the region near the peak of the distribution,
but also some part of the tail begins to approach a thermal
distribution. This indicates that for the central Au+Au collisions at
large $\snn$ the medium densities are high enough, resulting in
sufficient interactions among the constituents, to lead
to thermalization. The relaxation time scales for the system are 
comparable to or smaller than the life time of the fireball.

Our studies
shows a clear indication that an expanding
femto-scale QCD system formed in central collisions of high energy
heavy-ions~\cite{Adamczyk:2017iwn} attains thermalization. Furthermore, this test shows that neither synthetic data, nor peripheral collisions
can be considered as thermal or behaves like a bulk matter in thermodynamics. Interestingly the test also indicates that
central collisions at the two lowest $\snn$ may not be thermal. This
lack of thermalization can be due to the fireballs being too dilute to
ever come to equilibrium, or for there to be long relaxation time
at these energies. Further studies are needed to establish which is the
case, and thereby explore the part of the phase boundary which may be
probed by the low-energy experiments.

For a thermal distribution, all orders of the cumulants
should correspond to the same set of thermal parameters.
We have not included fourth and higher order
cumulants in this study. The higher order cumulants are, in principle,
more sensitive to correlation lengths, and therefore are more sensitive probes
of physics related to critical point and the nature of phase transition. 

\noindent{\bf Data availability}. The data that support the findings of this study are available from the corresponding 
author upon reasonable request.

\noindent{\bf Code availability}. The hadron resonance model code is available from the corresponding author upon reasonable request.

\noindent{\bf Acknowledgements} We thank Dr. Subhasis Samanta and
Mr. Ashish Pandav for discussions related to this work.



\begin{thebibliography}{10}
\expandafter\ifx\csname url\endcsname\relax
  \def\url#1{\texttt{#1}}\fi
\expandafter\ifx\csname urlprefix\endcsname\relax\def\urlprefix{URL }\fi
\providecommand{\bibinfo}[2]{#2}
\providecommand{\eprint}[2][]{\url{#2}}

\bibitem{white3} 
\bibinfo{author}{Adams,J.}\emph{et~al.},
\newblock \bibinfo{title}{{Experimental and theoretical challenges in the search for the quark gluon plasma: The STAR Collaboration's critical assessment of the evidence from RHIC collisions}}, 
\newblock \emph{\bibinfo{journal}{Nucl. Phys.}}
  \textbf{\bibinfo{volume}{A757}}, \bibinfo{pages}{102}
  (\bibinfo{year}{2005}). 

\bibitem{white4} 
\bibinfo{author}{Adcox,K.}\emph{et~al.},
\newblock \bibinfo{title}{{Formation of dense partonic matter in relativistic nucleus-nucleus collisions at RHIC: Experimental evaluation by the PHENIX collaboration}}, 
\newblock \emph{\bibinfo{journal}{Nucl. Phys.}}
  \textbf{\bibinfo{volume}{A757}}, \bibinfo{pages}{184}
  (\bibinfo{year}{2005}). 


\bibitem{Aamodt:2010pa} 
\bibinfo{author}{Aamodt,K.}\emph{et~al.},
\newblock \bibinfo{title}{{Elliptic flow of charged particles in Pb-Pb collisions at 2.76 TeV}}, 
\newblock \emph{\bibinfo{journal}{Phys. Rev. Lett.}}
  \textbf{\bibinfo{volume}{105}}, \bibinfo{pages}{252302}
  (\bibinfo{year}{2010}). 

 \bibitem{Aamodt:2010jd} 
\bibinfo{author}{Aamodt,K.}\emph{et~al.},
\newblock \bibinfo{title}{{Suppression of Charged Particle Production at Large Transverse Momentum in Central Pb-Pb Collisions at $\sqrt{s_{NN}} =$ 2.76 TeV}}, 
\newblock \emph{\bibinfo{journal}{Phys. Lett.}}
  \textbf{\bibinfo{volume}{B696}}, \bibinfo{pages}{30}
  (\bibinfo{year}{2011}).

 \bibitem{Chatrchyan:2011sx} 
\bibinfo{author}{Chatrchyan,S.}\emph{et~al.},
\newblock \bibinfo{title}{{Observation and studies of jet quenching in PbPb collisions at nucleon-nucleon center-of-mass energy = 2.76 TeV}}, 
\newblock \emph{\bibinfo{journal}{Phys. Rev.}}
  \textbf{\bibinfo{volume}{C84}}, \bibinfo{pages}{024906}
  (\bibinfo{year}{2011}). 


\bibitem{Gyulassy:2004zy} 
\bibinfo{author}{Gyulassy,~M. and McLerran,~L.D.}, 
\newblock \bibinfo{title}{{New forms of QCD matter discovered at RHIC}}, 
\newblock \emph{\bibinfo{journal}{ Nucl. Phys.}}
  \textbf{\bibinfo{volume}{A750}}, \bibinfo{pages}{30}
  (\bibinfo{year}{2005}).

  
\bibitem{Lattimer:2004pg} 
\bibinfo{author}{Lattimer,~J.M. and Prakash,~M.}, 
\newblock \bibinfo{title}{{The physics of neutron stars}}, 
\newblock \emph{\bibinfo{journal}{Science}}
  \textbf{\bibinfo{volume}{304}}, \bibinfo{pages}{536}
  (\bibinfo{year}{2004}).

  \bibitem{Csernai:2006zz} 
\bibinfo{author}{Csernai,~L.P., Kapusta,~J.I. and Mclerran,~L.D.}, 
\newblock \bibinfo{title}{{On the Strongly-Interacting Low-Viscosity Matter Created in Relativistic Nuclear Collisions}}, 
\newblock \emph{\bibinfo{journal}{Phys. Rev. Lett.}}
  \textbf{\bibinfo{volume}{97}}, \bibinfo{pages}{152303}
  (\bibinfo{year}{2006}).

\bibitem{Romatschke:2007mq} 
\bibinfo{author}{Romatschke,~P. and Romatschke,~U.}, 
\newblock \bibinfo{title}{{Viscosity Information from Relativistic Nuclear Collisions: How Perfect is the Fluid Observed at RHIC?}}, 
\newblock \emph{\bibinfo{journal}{Phys. Rev. Lett.}}
  \textbf{\bibinfo{volume}{99}}, \bibinfo{pages}{172301}
  (\bibinfo{year}{2007}).

 \bibitem{Kovtun:2004de} 
\bibinfo{author}{Kovtun,~P., Son,D.T. and Starinets,~A.O.}, 
\newblock \bibinfo{title}{{Viscosity in strongly interacting quantum field theories from black hole physics}}, 
\newblock \emph{\bibinfo{journal}{Phys. Rev. Lett.}}
  \textbf{\bibinfo{volume}{94}}, \bibinfo{pages}{111601}
  (\bibinfo{year}{2005}). 


 \bibitem{STAR:2017ckg} 
\bibinfo{author}{Adamczyk,~L..}\emph{et~al.},
\newblock \bibinfo{title}{{Global $\Lambda$ hyperon polarization in nuclear collisions: evidence for the most vortical fluid}}, 
\newblock \emph{\bibinfo{journal}{Nature}}
  \textbf{\bibinfo{volume}{548}}, \bibinfo{pages}{62}
  (\bibinfo{year}{2017}). 

\bibitem{Aoki:2006we} 
\bibinfo{author}{Aoki,~Y., Endrodi,~G., Fodor,~Z., Katz.~S.D. and Szabo,~K.K.}, 
\newblock \bibinfo{title}{{The Order of the quantum chromodynamics transition predicted by the standard model of particle physics}}, 
\newblock \emph{\bibinfo{journal}{Nature}}
  \textbf{\bibinfo{volume}{443}}, \bibinfo{pages}{675}
  (\bibinfo{year}{2006}).

   \bibitem{Gavai:2004sd} 
\bibinfo{author}{Gavai,R.V. and Gupta,~S.}, 
\newblock \bibinfo{title}{{The Critical end point of QCD}}, 
\newblock \emph{\bibinfo{journal}{Phys. Rev.}}
  \textbf{\bibinfo{volume}{D71}}, \bibinfo{pages}{114014}
  (\bibinfo{year}{2005}).

  
 \bibitem{Datta:2016ukp}
\bibinfo{author}{Datta,~S., Gavai,~R.V. and Gupta,~S.}, 
\newblock \bibinfo{title}{{Quark number susceptibilities and equation of state at finite chemical potential in staggered QCD with Nt=8}}, 
\newblock \emph{\bibinfo{journal}{Phys. Rev.}}
  \textbf{\bibinfo{volume}{D95}}, \bibinfo{pages}{054512}
  (\bibinfo{year}{2017}).

  \bibitem{DElia:2018fjp}
\bibinfo{author}{D'Elia,~M.}, 
\newblock \bibinfo{title}{{High-Temperature QCD: theory overview}}, 
\newblock \emph{\bibinfo{journal}{Nucl. Phys.}}
  \textbf{\bibinfo{volume}{A982}}, \bibinfo{pages}{99}
  (\bibinfo{year}{2019}).

  
 \bibitem{Stephanov:1999zu} 
\bibinfo{author}{Stephanov,~M.A., Rajagopal,~K. and Shuryak,~E.V.}, 
\newblock \bibinfo{title}{{Event-by-event fluctuations in heavy ion collisions and the QCD critical point}}, 
\newblock \emph{\bibinfo{journal}{Phys. Rev.}}
  \textbf{\bibinfo{volume}{D60}}, \bibinfo{pages}{114028}
  (\bibinfo{year}{1999}).

    
\bibitem{Hatta:2003wn} 
\bibinfo{author}{Hatta,~Y. and Stephanov,~M.A.}, 
\newblock \bibinfo{title}{{Proton number fluctuation as a signal of the QCD critical endpoint}}, 
\newblock \emph{\bibinfo{journal}{Phys. Rev. Lett.}}
  \textbf{\bibinfo{volume}{91}}, \bibinfo{pages}{102003}
  (\bibinfo{year}{2003}).

\bibitem{Stephanov:2008qz} 
\bibinfo{author}{Stephanov,~M.A.}, 
\newblock \bibinfo{title}{{Non-Gaussian fluctuations near the QCD critical point}}, 
\newblock \emph{\bibinfo{journal}{Phys. Rev. Lett.}}
  \textbf{\bibinfo{volume}{102}}, \bibinfo{pages}{032301}
  (\bibinfo{year}{2009}). 

 \bibitem{Asakawa:2009aj} 
\bibinfo{author}{Asakawa,~M., Ejiri,~S. and Kitazawa,~M.}, 
\newblock \bibinfo{title}{{Third moments of conserved charges as probes of QCD phase structure}}, 
\newblock \emph{\bibinfo{journal}{Phys. Rev. Lett.}}
  \textbf{\bibinfo{volume}{103}}, \bibinfo{pages}{262301}
  (\bibinfo{year}{2009}). 


\bibitem{Adamczyk:2017iwn} 
\bibinfo{author}{Adamczyk,~L..}\emph{et~al.}, 
\newblock \bibinfo{title}{{Bulk Properties of the Medium Produced in Relativistic Heavy-Ion Collisions from the Beam Energy Scan Program}}, 
\newblock \emph{\bibinfo{journal}{Phys. Rev.}}
  \textbf{\bibinfo{volume}{C96}}, \bibinfo{pages}{044904}
  (\bibinfo{year}{2017}). 

\bibitem{BraunMunzinger:2008tz} 
\bibinfo{author}{Braun-Munzinger,~P. and Wambach,~J.}, 
\newblock \bibinfo{title}{{The Phase Diagram of Strongly-Interacting Matter}}, 
\newblock \emph{\bibinfo{journal}{Rev. Mod. Phys.}}
  \textbf{\bibinfo{volume}{81}}, \bibinfo{pages}{1031}
  (\bibinfo{year}{2009}). 


 \bibitem{Fukushima:2010bq} 
\bibinfo{author}{Fukushima,~K. and Hatsuda,~T.}, 
\newblock \bibinfo{title}{{The phase diagram of dense QCD}}, 
\newblock \emph{\bibinfo{journal}{Rept. Prog. Phys.}}
  \textbf{\bibinfo{volume}{74}}, \bibinfo{pages}{014001}
  (\bibinfo{year}{2011}). 

\bibitem{Adamczyk:2013dal} 
\bibinfo{author}{Adamczyk,~L.}\emph{et~al.},
\newblock \bibinfo{title}{{Energy Dependence of Moments of Net-proton Multiplicity Distributions at RHIC}}, 
\newblock \emph{\bibinfo{journal}{Phys. Rev. Lett.}}
  \textbf{\bibinfo{volume}{112}}, \bibinfo{pages}{032302}
  (\bibinfo{year}{2014}).

 \bibitem{Aggarwal:2010wy} 
\bibinfo{author}{Aggarwal,~M.M.}\emph{et~al.},
\newblock \bibinfo{title}{{Higher Moments of Net-proton Multiplicity Distributions at RHIC}}, 
\newblock \emph{\bibinfo{journal}{Phys. Rev. Lett.}}
  \textbf{\bibinfo{volume}{110}}, \bibinfo{pages}{022302}
  (\bibinfo{year}{2010}).
  
 \bibitem{Luo:2015ewa} 
\bibinfo{author}{Luo,~X.}, 
\newblock \bibinfo{title}{{Energy Dependence of Moments of Net-Proton and Net-Charge Multiplicity Distributions at STAR}}, 
\newblock \emph{\bibinfo{journal}{PoS CPOD}}
  \textbf{\bibinfo{volume}{2014}}, \bibinfo{pages}{019}
  (\bibinfo{year}{2015}).

\bibitem{Smoot:1992td} 
\bibinfo{author}{Smoot,~G.F.}\emph{et~al.},
\newblock \bibinfo{title}{{Structure in the COBE differential microwave radiometer first year maps}}, 
\newblock \emph{\bibinfo{journal}{Astrophys. J.}}
  \textbf{\bibinfo{volume}{396}}, \bibinfo{pages}{L1}
  (\bibinfo{year}{1992}).

 \bibitem{Komatsu:2010fb} 
\bibinfo{author}{Komatsu,~E.}\emph{et~al.},
\newblock \bibinfo{title}{{Seven-Year Wilkinson Microwave Anisotropy Probe (WMAP) Observations: Cosmological Interpretation}}, 
\newblock \emph{\bibinfo{journal}{Astrophys. J. Suppl.}}
  \textbf{\bibinfo{volume}{192}}, \bibinfo{pages}{18}
  (\bibinfo{year}{2011}).

 \bibitem{Cleymans:2005xv} 
\bibinfo{author}{Cleymans,~J.,Oeschler,~H., Redlich,~K. and Wheaton,~S.}, 
\newblock \bibinfo{title}{{Comparison of chemical freeze-out criteria in heavy-ion collisions}}, 
\newblock \emph{\bibinfo{journal}{Phys. Rev.}}
  \textbf{\bibinfo{volume}{C73}}, \bibinfo{pages}{034905}
  (\bibinfo{year}{2006}).

 \bibitem{Andronic:2005yp} 
\bibinfo{author}{Andronic,~A., Braun-Munzinger,~P. and Stachel,~J.}, 
\newblock \bibinfo{title}{{Hadron production in central nucleus-nucleus collisions at chemical freeze-out}}, 
\newblock \emph{\bibinfo{journal}{Nucl. Phys.}}
  \textbf{\bibinfo{volume}{A772}}, \bibinfo{pages}{167}
  (\bibinfo{year}{2006}).
  
 \bibitem{Becattini:2003wp} 
\bibinfo{author}{Becattini,~F., Gazdzicki,~M., Keranen,~A., Manninen,~J. and Stock,~R.}, 
\newblock \bibinfo{title}{{Chemical equilibrium in nucleus nucleus collisions at relativistic energies}}, 
\newblock \emph{\bibinfo{journal}{Phys. Rev.}}
  \textbf{\bibinfo{volume}{C69}}, \bibinfo{pages}{024905}
  (\bibinfo{year}{2004}).

 \bibitem{Garg:2013ata} 
\bibinfo{author}{Garg,~P., Mishra,~D.K., Netrakanti,~P.K., Mohanty,~B., Mohanty,~A.K., 
  Singh,B.K., and Xu,~N.}, 
\newblock \bibinfo{title}{{Conserved number fluctuations in a hadron resonance gas model}}, 
\newblock \emph{\bibinfo{journal}{Phys. Lett.}}
  \textbf{\bibinfo{volume}{B726}}, \bibinfo{pages}{691}
  (\bibinfo{year}{2013}). 

  \bibitem{Andronic:2017pug} 
\bibinfo{author}{Andronic,~A., Braun-Munzinger,~P., Redlich,~K. and Stachel,~J.}, 
\newblock \bibinfo{title}{{Decoding the phase structure of QCD via particle production at high energy}}, 
\newblock \emph{\bibinfo{journal}{Nature}}
  \textbf{\bibinfo{volume}{561}}, \bibinfo{pages}{321}
  (\bibinfo{year}{2018}). 

\bibitem{Chatterjee:2013yga} 
\bibinfo{author}{Chatterjee,~S., Godbole,~R.M. and Gupta,~S.}, 
\newblock \bibinfo{title}{{Strange freezeout}}, 
\newblock \emph{\bibinfo{journal}{Phys. Lett.}}
  \textbf{\bibinfo{volume}{B727}}, \bibinfo{pages}{554}
  (\bibinfo{year}{2013}).

\bibitem{Venugopalan:1992hy} 
\bibinfo{author}{Venugopalan,~R. and Prakash,~M.}, 
\newblock \bibinfo{title}{{Thermal properties of interacting hadrons}}, 
\newblock \emph{\bibinfo{journal}{Nucl. Phys.}}
  \textbf{\bibinfo{volume}{A546}}, \bibinfo{pages}{718}
  (\bibinfo{year}{1992}).


 \bibitem{Yen:1997rv} 
\bibinfo{author}{Yen,~G.D., Gorenstein,~M.I., Greiner,~W. and Yang,~S.N.}, 
\newblock \bibinfo{title}{{Excluded volume hadron gas model for particle number ratios in A+A collisions}}, 
\newblock \emph{\bibinfo{journal}{Phys. Rev.}}
  \textbf{\bibinfo{volume}{C56}}, \bibinfo{pages}{2210}
  (\bibinfo{year}{1997}). 

\bibitem{Chatterjee:2009km} 
\bibinfo{author}{Chatterjee,~S., Godbole,~R.M. and Gupta,~S.}, 
\newblock \bibinfo{title}{{Stabilizing hadron resonance gas models}}, 
\newblock \emph{\bibinfo{journal}{Phys. Rev.}}
  \textbf{\bibinfo{volume}{C81}}, \bibinfo{pages}{044907}
  (\bibinfo{year}{2010}).

 \bibitem{Vovchenko:2016rkn} 
\bibinfo{author}{Vovchenko,~V., Gorenstein,~M.I. and Stoecker,~H.}, 
\newblock \bibinfo{title}{{van der Waals Interactions in Hadron Resonance Gas: From Nuclear Matter to Lattice QCD}}, 
\newblock \emph{\bibinfo{journal}{Phys. Rev. Lett.}}
  \textbf{\bibinfo{volume}{118}}, \bibinfo{pages}{182301}
  (\bibinfo{year}{2017}).

 \bibitem{Samanta:2017yhh} 
\bibinfo{author}{Samanta,~S. and Mohanty,~B.}, 
\newblock \bibinfo{title}{{Criticality in a Hadron Resonance Gas model with the van der Waals interaction}}, 
\newblock \emph{\bibinfo{journal}{Phys. Rev.}}
  \textbf{\bibinfo{volume}{C97}}, \bibinfo{pages}{015201}
  (\bibinfo{year}{2018}).
  
 \bibitem{Chatterjee:2017yhp} 
\bibinfo{author}{Chatterjee,~S., Mishra,~D., Mohanty,~B. and Samanta,~S.}, 
\newblock \bibinfo{title}{{Freezeout systematics due to the hadron spectrum}}, 
\newblock \emph{\bibinfo{journal}{Phys. Rev.}}
  \textbf{\bibinfo{volume}{C96}}, \bibinfo{pages}{054907}
  (\bibinfo{year}{2017}).

 \bibitem{Becattini:1995if} 
\bibinfo{author}{Becattini,~F.}, 
\newblock \bibinfo{title}{{A Thermodynamical approach to hadron production in e+ e- collisions}}, 
\newblock \emph{\bibinfo{journal}{Z. Phys.}}
  \textbf{\bibinfo{volume}{C69}}, \bibinfo{pages}{485}
  (\bibinfo{year}{1996}).

\bibitem{Becattini:1997rv} 
\bibinfo{author}{Becattini,~F. and Heinz,~U.W.}, 
\newblock \bibinfo{title}{{Thermal hadron production in p p and p anti-p collisions}}, 
\newblock \emph{\bibinfo{journal}{Z. Phys.}}
  \textbf{\bibinfo{volume}{C76}}, \bibinfo{pages}{269}
  (\bibinfo{year}{1997}). 

\bibitem{Das:2016muc} 
\bibinfo{author}{Das,~S., Mishra,~D., Chatterjee,~S. and Mohanty,~B.}, 
\newblock \bibinfo{title}{{Freeze-out conditions in proton-proton collisions at the highest energies available at the BNL Relativistic Heavy Ion Collider and the CERN Large Hadron Collider}}, 
\newblock \emph{\bibinfo{journal}{Phys. Rev.}}
\textbf{\bibinfo{volume}{C95}}, \bibinfo{pages}{014912}
  (\bibinfo{year}{2017}).

\bibitem{BraunMunzinger:2007zz} 
\bibinfo{author}{Braun-Munzinger,~P. and Stachel,~J.}, 
\newblock \bibinfo{title}{{The quest for the quark-gluon plasma}}, 
\newblock \emph{\bibinfo{journal}{Nature}}
  \textbf{\bibinfo{volume}{448}}, \bibinfo{pages}{302}
  (\bibinfo{year}{2007}). 


  \bibitem{Becattini:2012xb} 
\bibinfo{author}{Becattini,~F., Bleicher,~M., Kollegger,~T., Schuster,~T., Steinheimer,~J. and Stock,~R.}
\newblock \bibinfo{title}{{Hadron Formation in Relativistic Nuclear Collisions and the QCD Phase Diagram}}, 
\newblock \emph{\bibinfo{journal}{Phys. Rev. Lett.}}
  \textbf{\bibinfo{volume}{111}}, \bibinfo{pages}{082302}
  (\bibinfo{year}{2013}). 

 \bibitem{Gavai:2010zn} 
\bibinfo{author}{Gavai,~R.V. and Gupta,~S.}, 
\newblock \bibinfo{title}{{Lattice QCD predictions for shapes of event distributions along the freezeout curve in heavy-ion collisions}}, 
\newblock \emph{\bibinfo{journal}{Phys. Lett.}}
  \textbf{\bibinfo{volume}{B696}}, \bibinfo{pages}{459}
  (\bibinfo{year}{2011}). 

 \bibitem{Bazavov:2012vg} 
\bibinfo{author}{Bazavov,~A.}\emph{et~al.}, 
\newblock \bibinfo{title}{{Freeze-out Conditions in Heavy Ion Collisions from QCD Thermodynamics}}, 
\newblock \emph{\bibinfo{journal}{Phys. Rev. Lett.}}
  \textbf{\bibinfo{volume}{109}}, \bibinfo{pages}{192302}
  (\bibinfo{year}{2012}). 

  \bibitem{Borsanyi:2014ewa} 
\bibinfo{author}{Borsanyi,~S., Fodor,~Z., Katz,~S.~D., Krieg,~S., Ratti,~C. and Szabo,~K.~K.}
\newblock \bibinfo{title}{{Freeze-out parameters from electric charge and baryon number fluctuations: is there consistency?}}, 
\newblock \emph{\bibinfo{journal}{Phys. Rev. Lett.}}
  \textbf{\bibinfo{volume}{113}}, \bibinfo{pages}{052301}
  (\bibinfo{year}{2014}). 

\bibitem{Bzdak:2012an} 
\bibinfo{author}{Bzdak,~A., Koch,~V., and Skokov,~V.}, 
\newblock \bibinfo{title}{{Baryon number conservation and the cumulants of the net proton distribution}}, 
\newblock \emph{\bibinfo{journal}{Phys. Rev.}}
\textbf{\bibinfo{volume}{C87}}, \bibinfo{pages}{014901}
  (\bibinfo{year}{2013}).

   \bibitem{Gupta:2009mu}
\bibinfo{author}{Gupta,~S.}, 
\newblock \bibinfo{title}{{Finding the critical end point of QCD: Lattice and experiment}}, 
\newblock \emph{\bibinfo{journal}{PoS CPOD}}
  \textbf{\bibinfo{volume}{2009}}, \bibinfo{pages}{025}
  (\bibinfo{year}{2009}).

\bibitem{Gupta:2011wh}
\bibinfo{author}{Gupta,~S., Luo,~X., Mohanty,~B., Ritter,~H.G. and Xu,~N.}, 
\newblock \bibinfo{title}{{Scale for the Phase Diagram of Quantum Chromodynamics}}, 
\newblock \emph{\bibinfo{journal}{Science}}
  \textbf{\bibinfo{volume}{332}}, \bibinfo{pages}{1525}
  (\bibinfo{year}{2011}).
  
\bibitem{Cheng:2008zh} 
\bibinfo{author}{Cheng,~M.}\emph{et~al.},
\newblock \bibinfo{title}{{Baryon Number, Strangeness and Electric Charge Fluctuations in QCD at High Temperature}}, 
\newblock \emph{\bibinfo{journal}{Phys. Rev.}}
  \textbf{\bibinfo{volume}{D79}}, \bibinfo{pages}{074505}
  (\bibinfo{year}{2009}).

\bibitem{Bellwied:2013cta} 
\bibinfo{author}{Bellwied,~R., Borsanyi,~S., Fodor,~Z., Katz~S.D. and Ratti,~C.}
\newblock \bibinfo{title}{{Is there a flavor hierarchy in the deconfinement transition of QCD?}}, 
\newblock \emph{\bibinfo{journal}{Phys. Rev. Lett.}}
  \textbf{\bibinfo{volume}{111}}, \bibinfo{pages}{202302}
  (\bibinfo{year}{2013}). 

 \bibitem{Adamczyk:2017wsl} 
\bibinfo{author}{Adamczyk,~L.}\emph{et~al.}, 
\newblock \bibinfo{title}{{Collision Energy Dependence of Moments of Net-Kaon Multiplicity Distributions at RHIC}}, 
\newblock \emph{\bibinfo{journal}{Phys. Lett.}}
  \textbf{\bibinfo{volume}{B785}}, \bibinfo{pages}{551}
  (\bibinfo{year}{2018}). 

\bibitem{Adamczyk:2014fia} 
\bibinfo{author}{Adamczyk,~L.}\emph{et~al.}, 
\newblock \bibinfo{title}{{Beam energy dependence of moments of the net-charge multiplicity 
  distributions in Au+Au collisions at RHIC}}, 
\newblock \emph{\bibinfo{journal}{Phys. Rev. Lett.}}
  \textbf{\bibinfo{volume}{113}}, \bibinfo{pages}{092301}
  (\bibinfo{year}{2014}). 

\bibitem{Adam:2019xmk} 
\bibinfo{author}{Adamczyk,~L..}\emph{et~al.}, 
\newblock \bibinfo{title}{{Collision-energy dependence of second-order off-diagonal and diagonal cumulants of net-charge, net-proton, and net-kaon multiplicity distributions in Au + Au collisions}}, 
\newblock \emph{\bibinfo{journal}{Phys. Rev.}}
  \textbf{\bibinfo{volume}{C100}}, \bibinfo{pages}{014902}
  (\bibinfo{year}{2019}).

\bibitem{Kitazawa:2012at} 
\bibinfo{author}{Kitazawa,~M. and Asakawa,~M.}, 
\newblock \bibinfo{title}{{Relation between baryon number fluctuations and experimentally observed proton number fluctuations in relativistic heavy ion collision}}, 
\newblock \emph{\bibinfo{journal}{Phys. Rev.}}
  \textbf{\bibinfo{volume}{C86}}, \bibinfo{pages}{024904}
  (\bibinfo{year}{2011}).

\bibitem{Zhou:2017jfk} 
\bibinfo{author}{Zhou,~C., Xu,~J., Luo,~X. and Liu,~F.}, 
\newblock \bibinfo{title}{{Cumulants of event-by-event net-strangeness distributions in Au+Au collisions at $\sqrt{s_\mathrm{NN}}$=7.7-200 GeV from UrQMD model}}, 
\newblock \emph{\bibinfo{journal}{Phys. Rev.}}
  \textbf{\bibinfo{volume}{C96}}, \bibinfo{pages}{014909}
  (\bibinfo{year}{2017}). 

 \bibitem{Mishra:2016qyj}
\bibinfo{author}{Mishra,~D.K, Garg,~P., Netrakanti,~P.K. and Mohanty,~A.K.}, 
\newblock \bibinfo{title}{{Effect of resonance decay on conserved number fluctuations in a hadron resonance gas model}}, 
\newblock \emph{\bibinfo{journal}{Phys. Rev.}}
  \textbf{\bibinfo{volume}{C94}}, \bibinfo{pages}{014905}
  (\bibinfo{year}{2016}). 

 \bibitem{Nahrgang:2014fza}
\bibinfo{author}{Nahrgang,~M., Bluhm,~M., Alba,~P., Bellwied,~R. and Ratti,~C.}, 
\newblock \bibinfo{title}{{Impact of resonance regeneration and decay on the net-proton fluctuations in a hadron resonance gas}}, 
\newblock \emph{\bibinfo{journal}{Eur. Phys.}}
  \textbf{\bibinfo{volume}{C75}}, \bibinfo{pages}{573}
  (\bibinfo{year}{2015}). 

 \bibitem{Gavai:2005yk}
\bibinfo{author}{Gavai,~R.~V. and Gupta,~S.}, 
\newblock \bibinfo{title}{{Fluctuations, strangeness and quasi-quarks in heavy-ion collisions from lattice QCD}}, 
\newblock \emph{\bibinfo{journal}{Phys. Rev.}}
  \textbf{\bibinfo{volume}{D73}}, \bibinfo{pages}{014004}
  (\bibinfo{year}{2006}). 
  
 \bibitem{Alba:2014eba}
\bibinfo{author}{Alba,~P., Alberico,~W., Bellwied,~R., Bluhm,~M., Mantovani Sarti,~V., Nahrgang,~M. and Ratti,~C.}, 
\newblock \bibinfo{title}{{Freeze-out conditions from net-proton and net-charge fluctuations at RHIC}}, 
\newblock \emph{\bibinfo{journal}{Phys. Lett.}}
  \textbf{\bibinfo{volume}{B738}}, \bibinfo{pages}{305}
  (\bibinfo{year}{2014}). 

  \bibitem{Bleicher:1999xi}
\bibinfo{author}{Bleicher,~M.}\emph{et~al.}, 
\newblock \bibinfo{title}{{Relativistic hadron hadron collisions in the ultrarelativistic quantum molecular dynamics model}}, 
\newblock \emph{\bibinfo{journal}{J. Phys.}}
  \textbf{\bibinfo{volume}{G25}}, \bibinfo{pages}{1859}
  (\bibinfo{year}{1999}).

  \bibitem{Petersen:2008kb}
\bibinfo{author}{Petersen,~H., Bleicher,~M., Bass,~S.~A. and Stocker,~H.}, 
\newblock \bibinfo{title}{{UrQMD v2.3: Changes and Comparisons}}, 
\newblock \emph{\bibinfo{journal}{arXiv:0805.0567}}
(\bibinfo{year}{2008}).

  \bibitem{Lin:2004en}
\bibinfo{author}{Lin,~Z.~W., Ko,~C.~M., Li,~B.~A., Zhang,~B. and Pal,~S.}, 
\newblock \bibinfo{title}{{A Multi-phase transport model for relativistic heavy ion collisions}}, 
\newblock \emph{\bibinfo{journal}{Phys. Rev.}}
 \textbf{\bibinfo{volume}{C72}}, \bibinfo{pages}{064901}
  (\bibinfo{year}{2005}).  

\end{thebibliography}
\end{document}